\def\gapprox{\lower.4ex\hbox{$\;\buildrel >\over{\scriptstyle\sim}\;$}}
\def\lapprox{\lower.4ex\hbox{$\;\buildrel <\over{\scriptstyle\sim}\;$}}

\def\ref#1   {\par\noindent\hangindent1cm {#1}}

 \documentstyle[aaspp4]{article}		
 \input epsf
\slugcomment{}
\lefthead{ASCHWANDEN}
\righthead{ELECTRON VERSUS PROTON TIMING}

\begin{document}
{{\sl THE ASTROPHYSICAL JOURNAL LETTERS}\hfill{Accepted, 1996 July 29}}

\title{		Electron versus Proton Timing Delays\\ 
		in Solar Flares}

\author{        Markus J. Aschwanden}
\affil{         Department of Astronomy, University of Maryland,
		College Park, MD 20742\\ 
		markus@astro.umd.edu}

\begin{abstract}
Both electrons and ions are accelerated in solar flares and carry nonthermal
energy from the acceleration site to the chromospheric energy loss site, but
the relative amount of energy carried by electrons versus ions is subject
of debate. In this {\sl Letter} we test whether the observed energy-dependent 
timing delays of 20-200 keV HXR emission can be explained in terms of 
propagating electrons versus protons. For a typical flare, we show that the 
timing delays of fast ($\lapprox 1$ s) {\sl HXR pulses} is consistent with 
time-of-flight differences of directly precipitating electrons, while the 
timing delays of the {\sl smooth HXR}
flux is consistent with collisional deflection times of trapped electrons.
We show that these HXR timing delays cannot be explained either by $\le 1$ MeV 
protons (as proposed in a model by Simnett \& Haines 1990),
because of their longer propagation and trapping times, or by $\approx 40$ MeV
protons (which have the same velocity as $\approx 20$ keV electrons), 
because of their longer trapping 
times and the excessive fluxes required to generate the HXRs. 
Thus, the HXR timing results clearly rule out protons as the primary 
generators of $\ge 20$ keV HXR emission.

\end{abstract}


\section{ 	INTRODUCTION 						}

Whether protons or electrons dominate particle acceleration and non-thermal
energy transport in solar flares is still subject of recent debate [e.g.
see {\sl Great Debate} articles by Cargill (1996), Simnett (1996), and Emslie
(1996)]. Moreover, proponents of proton-dominated models argue that even
the $\ge 20$ keV HXR bremsstrahlung is produced either by protons directly
(Heristchi 1986; 1987) or by proton-energized electrons (Simnett \& Haines
1990). The purpose of this {\sl Letter} is to complement these ongoing
debates with powerful new arguments that are supported by observations 
of energy-dependent hard X-ray (HXR) timing delays in solar flares.

Electrons have been favored for the interpretation of impulsive $\ge 20$ keV
HXRs because the inversion of the HXR spectrum (Brown 1971) in the framework 
of energy loss by collisional bremsstrahlung (in a relatively cold target)
could easily reproduce: (1) an electron spectrum that contains the necessary and
sufficient energy content to explain the non-thermal HXR emission,
(2) the gyrosynchrotron emission observed at microwave wavelengths, and (3) 
the soft X-ray emission resulting from the heated, upflowing chromospheric
plasma. More detailed arguments for the case of electron dominance in solar
flare energy transport can be found in Emslie (1996), Emslie et al. (1996), 
and references therein.
                              
Protons, on the other hand, have been invoked as primary energy carriers
(Simnett 1986; 1995) mainly because it minimizes the number of energy-carrying 
particles (thanks to their intrinsically higher rest mass, i.e. 
$m_p/m_e = 1836$), which in turn also minimizes 
return currents (at least for high-energy protons with $E \gg 20$ keV, 
see, e.g., Simnett \& Strong 1984, Emslie \& Brown 1985). Observational support
for the proton hypothesis was mainly gathered by indirect effects, e.g.
(1) acceleration mechanisms (e.g. shocks) can have greater efficiencies
for protons because of their larger gyroradii, (2) the proton spectrum 
at energies $\le 10$ MeV could be considerably steeper than previously 
supposed (Ramaty et al. 1995), based on recent measurements of Share 
\& Murphy (1995), (3) detection of interplanetary high-energy ($\ge 40$ MeV) 
protons (Evenson et al. 1984), etc. A discussion of arguments 
in favor of protons can be found in a recent review by Simnett (1995),
and further arguments of proton versus electron beam models are discussed
in Brown et al. (1990). 

A new aspect that we would like to contribute to the electron-versus-proton 
debate is based on new timing results from HXR observations in
solar flares, analyzed in a series of recent studies
(Aschwanden, Schwartz \& Alt 1995; Aschwanden \& Schwartz 1995; 1996; 
Aschwanden 1996; Aschwanden et al. 1996a; 1996b; 1996c). The energy-dependent
timing delays $\tau (E)$ of fast ($\lapprox 1$ s) {\sl HXR pulses} during 
the impulsive flare phase have been found to be consistent with
electron time-of-flight differences, while the timing delays 
of the {\sl smooth HXR} flux are comparable with electron
trapping time scales. The interpretation of hard X-ray time structures in
terms of these two components (associated with directly-precipitating versus 
trapped particles) is a basic assumption used in the following
discussion, in contrast to the widely-held hypothesis that the 
lower envelope of the rapidly-varying hard X-ray flux consists of an unresolved
pileup of elementary pulses with similar characteristics as the rapid
fluctuations (e.g. Kiplinger et al. 1983, 1984; Machado et al. 1993). 
While protons have not been considered in our previous 
studies, we would like to explore in this {\sl Letter} whether proton theories
could be reconciled with the observed
HXR timing delays, and what the required proton
energies would be. Because the timing delays of HXRs are directly related to the
particle velocities, timing information provides a direct means of 
constraining the kinetic energy and rest mass of the involved particles.

In Section \S 2 we compare theoretical electron and proton time-of-flight 
differences 
with the observed delays of fast {\sl HXR pulses}. 
In Section \S 3 we discuss theoretical trapping times
in solar flare loops and compare them with the observed
delays of the {\sl smooth HXR} emission. 
Section \S 4 contains conclusions on the question of which 
species of particles is more likely to reproduce the observed HXR timing delays.

\section{ Particle Time-of-Flight Differences }

Both proponents of electron-dominated and proton-dominated flare models
agree with the assumption that particles are accelerated in the coronal part
of a flare loop and that the HXRs observed from the loop footpoints
are produced by collisional bremsstrahlung in the chromosphere. 
The main difference between the two models is
the question of the particle species that carries the energy from the
acceleration site to the energy loss site. In the case of proton-dominated
models, either the protons produce HXR's directly by inverse bremsstrahlung
(Emslie \& Brown 1985; Heristchi 1986; 1987), or a neutral beam is assumed 
to precipitate to the chromosphere, where an electric field arising from charge
separation accelerates HXR-emitting electrons (Simnett \& Haines 1990). 
In the latter model, electrons are thought to be stripped by Coulomb scattering
from a neutralized ion beam as it impacts the chromosphere. This charge
separation process may set up an electric field that is able to produce
runaway acceleration of electrons responsible for the $\ge 20$ keV
HXR bremsstrahlung. This model has been severely criticized by Emslie (1996).

Thus, the timing of the two models can be tested in first order by comparing
the particle time-of-flight differences occurring during the propagation from
the coronal acceleration site to the chromospheric energy loss site. If the
energy-dependent timing delays 
of the HXRs is dominated by such propagation effects,
the HXR emission produced by the low-energy particles should always be delayed
with respect to the high-energy particles, because of their lower velocity.
Energy loss times and acceleration times seem not to dominate the timing 
delays of
($\lapprox 1$ s) HXR pulses in most of the flares, because they do not 
reproduce the correct sign in the energy-dependent timing delays
(Aschwanden et al. 1996a; Aschwanden 1996).
Thus, the time-of-flight difference ${\tau}_{ij}$ between two particles 
$i$ and $j$ arriving at 
the chromospheric energy loss-site is determined by their velocities 
${\beta}_i=v_i/c$ (or ${\beta}_j=v_j/c$) and the path length $l$,
\begin{equation}
	{\tau}_{ij} = { l \over c } ({1 \over {\beta}_i} - {1 \over {\beta}_j})
	\ .
\end{equation}
The relativistic velocity ${\beta}$ is defined by the kinetic energy
$E_{kin}=m c^2 ({\gamma} - 1)$ of the particles and the Lorentz factor
${\gamma} = 1 / \sqrt{1 - {\beta}^2}$, i.e.
\begin{equation}
	\beta(E_{kin}) = \sqrt{ 1 - \big[ {E_{kin} \over m c^2} + 1 \big]^{-2}}
\end{equation}
If we compare electrons with a kinetic energy $E_e$ and rest mass $m_e$
with protons of kinetic energy $E_p$ and rest mass $m_p$, the 
velocity ratio is
\begin{equation}
	{{\beta}_p \over {\beta}_e} =
	{\sqrt{ 1 - [{E_p \over m_p c^2} + 1]^{-2}} \over 
	 \sqrt{ 1 - [{E_e \over m_e c^2} + 1]^{-2}}} 
	\approx \sqrt{ {E_p \over E_e} \cdot {m_e \over m_p}}
	= {1 \over 43} \sqrt{E_p \over E_e} \ ,
\end{equation}
where the right-hand approximation applies to the non-relativistic (or mildly
relativistic) case. For instance, to obtain identical velocities for electrons
and protons, the kinetic energy of the protons has to be a factor of
$m_p/m_e=1836$ larger, e.g. 40 MeV protons have the same velocity as 20 keV
electrons.

In the case of protons, it was proposed that flare HXRs could be produced
by an "inverse bremsstrahlung process" (Boldt and Serlemitsos 1969; Emslie
\& Brown 1985; Heristchi 1986; 1987). Protons would produce HXR photons by 
collisional bremsstrahlung in collisional encounters with electrons. The 
relative velocity of a proton of energy
\begin{equation}
	E_p = ({m_p \over m_e}) E_e
\end{equation}
in an encounter with an ambient electron is the same as that of an electron of
energy $E_e$ in an encounter with a stationary proton. Since $\ge 40$ MeV 
protons have identical velocities as $\ge 20$ keV electrons, they would also 
show the same time-of-flight differences as observed. 
However, the major problem with $\ge 40$ MeV protons is that the required
proton flux to produce the observed $\ge 20$ keV HXR flux would exceed the
proton flux deduced from nuclear gamma-ray line observations (at 2-10 MeV)
by a factor of $\approx 10^3$ (Ramaty 1985).

To alleviate the problem of $40$ MeV protons, Simnett \& Haines (1990)
proposed a neutral beam model, that requires only 0.1-1 MeV protons to
produce the observed $\ge 20$ keV HXR flux. After a neutral beam arrives
at the chromosphere, HXR emission is produced by secondary electrons
accelerated in the electric field that arises from the charge separation
of penetrating ions and thermalized primary electrons. 
Thus, the timing delays of
HXR emission are mainly determined by the coronal propagation time of
the 0.1-1 MeV protons in the neutral beam. However, these lower proton
energies correspond to velocities that would produce far greater
time-of-flight differences than for 40 MeV protons or 20 keV electrons.
According to Eq.3, the velocity of 1 MeV protons would be a factor of
${\beta}_p/{\beta}_e\le (1/43)\sqrt{1/0.02}=0.1$ smaller than for 20 keV
electrons, and thus, the time-of-flight differences would be 10 times
larger than for 20 keV electrons or 40 MeV protons. 

Fig.1 shows typical HXR time delay measurements in the energy range of 
40-300 keV for a major flare. The {\sl pulsed HXR} flux (Fig.1 bottom right)
exhibits time delays of $\tau (E) = t(40$ keV)$ - t(E) \lapprox 50$ ms.
These delays show the correct sign as expected
for particle time-of-flight differences, i.e. the lower energies lag the
high energies. Also the functional form of the energy-dependent time
delay is consistent with the expected model (Eq.1, thick line on
right-hand side bottom in Fig.1), within a high accuracy of $\lapprox 10$
ms. For the conversion of electron energies $E$ into HXR photon
energies $\varepsilon$, a conversion factor of $q_E=E/\varepsilon \approx 2$
was determined (see second line on right-hand side from bottom in Fig.1,
according to calculations in Aschwanden \& Schwartz 1996). 
Moreover, the time-of-flight distance, i.e. $l = 20,000$ km, 
is found to be close to a loop half length ($s = 14,000$ km), 
if electron velocities are employed. This ratio $l/s \approx 1.4$ was 
established to be scale-invariant for a
comprehensive set of loop sizes in different flares (Aschwanden et al.
1996b, 1996c).

If we employ protons to reproduce these observed HXR delays, we can either
adjust the energy range or the time-of-flight distance. If we adjust
the energy range, protons with energies of $\ge 40$ MeV would be required 
to produce matching velocities. If we employ
1 MeV protons, the velocities would be 10 times smaller, thus
requiring a time-of-flight distance 10 times shorter to satisfy the same
timing delays (Eq.1; also depicted in Fig.1 middle right). 
Based on the statistics of flare loop sizes
(typically with loop radii of $\lapprox 20,000$ km, Aschwanden et
al. 1996c), a time-of-flight distance 10 times smaller ($l\lapprox
2000$ km) would be comparable with the height of the chromosphere,
requiring an acceleration site located near the transition region. Such
a scenario would have difficulty to explain the observed simultaneity 
of HXR emission in conjugate loop footpoints, which were found to be 
coincident within $\lapprox 0.1$ s (Sakao 1994; Kosugi 1996), but are
separated by considerably larger electron travel times of 0.5-1.0 s. 

Therefore, it seems that all proton models face some fundamental difficulties 
to reconcile the observed timing delays of impulsive HXR emission.  

\section{ Particle Trapping Times }

The bulk of HXR emission ($\approx 50-90\%$, depending on the modulation
depth of HXR pulses) is contained in smooth,
slowly-varying time structures, which moreover exhibit an opposite
sign of the energy-dependent time delay, compared with the
fast-fluctuating HXR pulses (Aschwanden et al. 1996b;
1996c). The timing delays of this {\sl smooth HXR} component was therefore
interpreted in terms of particle trapping or energy loss inside the
flare loop. This interpretation is mainly motivated by two reasons:
(1) trapping is expected to smear out rapid fluctuations of the acceleration
and injection process, and (2) the collisional deflection time 
or collisional energy loss time shows a functional dependence ($\tau
\propto E^{3/2}$) that reproduces the correct sign and fits the functional form 
of the observed delay ${\tau}(E)$ (see example in Fig.1, bottom left).

Let us consider the collisional deflection time, which in any case
should represent an upper limit of the particle trapping time in a flare
loop. The collisional deflection time is defined by (e.g. Benz 1993),
\begin{equation}
	t^{defl}(E) := {v^2 \over <\Delta v_{\perp}^2 / \Delta t >}
	= 9.5 \cdot 10^7 {E_{keV}^{3/2} \over n_e}
	({m_T \over m_e})^2 {1 \over Z^2_T}
	({20 \over \ln \Lambda}) \qquad [s]
\end{equation}
with $E_{keV}$ the kinetic energy, $m_T$ the rest mass, and $Z_T$ the
charge of the test particles, $n_e$ the electron density of the ambient plasma, 
and $\ln \Lambda$ the Coulomb logarithm. 
If we employ $50-100$ keV electrons (to produce $25-50$ keV HXR photons),
the expected deflection time is about $t^{defl}_e \approx 0.3-1.0$ s in a
flare loop with a typical electron density of $n_e \approx 10^{11}$
cm$^{-3}$. Considering the fit of $t^{defl}(E)$ to the observed time delay 
for the flare shown in Fig.1 (bottom left), we find that the {\sl smooth HXR} 
flux has timing delays that are consistent with a trapping model in terms 
of collisional deflection.
If we employ protons, the collisional deflection times would be a factor
of $(E_p/E_e)^{3/2} (m_p/m_e)^2 \approx 10^9-10^{11}$ longer
than for electrons, which is far beyond any delay ever measured in
HXRs or gamma rays. The collisional deflection time, though an upper
limit, is therefore no reasonable estimator of trapping times for
protons. Pitch-angle scattering by strong wave turbulence or cross-field
diffusion times in twisted magnetic flux tubes are believed to
restrict the trapping times of ions or protons in flare loops to time scales
of $\approx 10^2-10^4$ s (Ramaty \& Mandzhavidze 1994), which are a
factor of $\approx 10-10^3$ longer than collisional deflection times of 
electrons ($t^{defl}\approx 1-10$ s in typical flare loop densities; 
Aschwanden et al. 1996a; 1996b). 

If we employ protons to generate the $\ge 25$ keV HXR emission (e.g.
Simnett \& Haines 1990), the precipitating 
protons are necessarily required to have an identical timing as the observed 
HXRs. This requirement for the proton timing cannot be reconciled with 
the observed timing delays of the smooth HXR flux for the following reasons: 
(1) If the smooth HXR flux 
would be produced by directly precipitating protons, the proton timing would 
have an opposite sign because of time-of-flight differences; (2) If the
smooth HXR flux would be produced by initially trapped and subsequently
precipitating protons, the proton trapping times would be much longer
(a factor of $\approx 10-10^3$) than the observed HXR delays, which were
found to fit electron collisional deflection times (Fig.1, bottom left). 
In the latter case, it would also be very difficult
to understand that the trapping times of $>1$ MeV protons would have exactly
the same energy-dependence as the collisional deflection time of $\ge 20$ keV
electrons (see Fig.1, bottom left). 
Thus, we see no suitable possibility to adjust the proton timing 
to the observed HXR delays, neither in terms of time-of-flight differences
nor in terms of trapping time differences.

\section{ Conclusions }

From previous work we found that the energy-dependent timing delays ${\tau}(E)$
of solar flare HXR emission in the energy range of $\approx 20-200$ keV
are consistent with two timing processes: (1) the timing delays of the 
{\sl pulsed}
HXR flux are consistent with electron time-of-flight differences over
spatial scales that correspond roughly to half a loop length, and
(2) the timing delays of the 
{\sl smooth} HXR flux are consistent with collisional
deflection times for electrons in typical flare loop densities ($n_e \approx
10^{11}$ cm$^{-3}$).

In an attempt to explain the same timing results of {\sl HXR pulses} 
in terms of protons, we considered two energy ranges, which both face 
fundamental difficulties: (1) $\approx 40$ MeV
protons would match the same velocities and time-of-flight differences,
but would require a proton flux far in excess of gamma-ray line observations,
and (2) $\le 1$ MeV protons have a velocity $\approx 10$ times smaller than 
20 keV electrons and can only satisfy the observed time-of-flight differences
for flight distances $\approx 10$ times smaller, which are difficult to 
reconcile with the observed simultaneity of conjugate HXR footpoint emission. 

Employing protons to explain the observed timing delays of the {\sl smooth} HXR
flux would require that the trapping times of protons have the same energy 
dependence as collisional deflection times for electrons. This 
conflicts with estimates of proton trapping times, which are significantly
longer than those of electrons, in the case of collisional deflection
as well as for pitch-angle scattering by wave turbulence.

Based on these timing arguments it appears to be very unlikely that
protons are responsible for producing $\ge 20$ keV HXR bremsstrahlung, e.g.
as proposed by Simnett \& Haines (1990). However, these
timing results do not exclude that an arbitrary amount of ions (or protons) 
are accelerated concomitantly with electrons and propagate to the
chromosphere, as evidenced by the observed nuclear gamma-ray lines 
(at 1-10 MeV) in large flares. The main conclusion of this study is that
{\bf protons cannot be responsible for $\ge 20$ keV HXR emission},
because their timing would be inconsistent with the observed HXR delays.

\medskip
\acknowledgements
The author thanks to Gordon Emslie, Takeo Kosugi, and the anonymous referee 
for their thougthful comments.  The work was supported by SR\&T grants 
NAG-5-2352 and NAGW-5078.

\clearpage

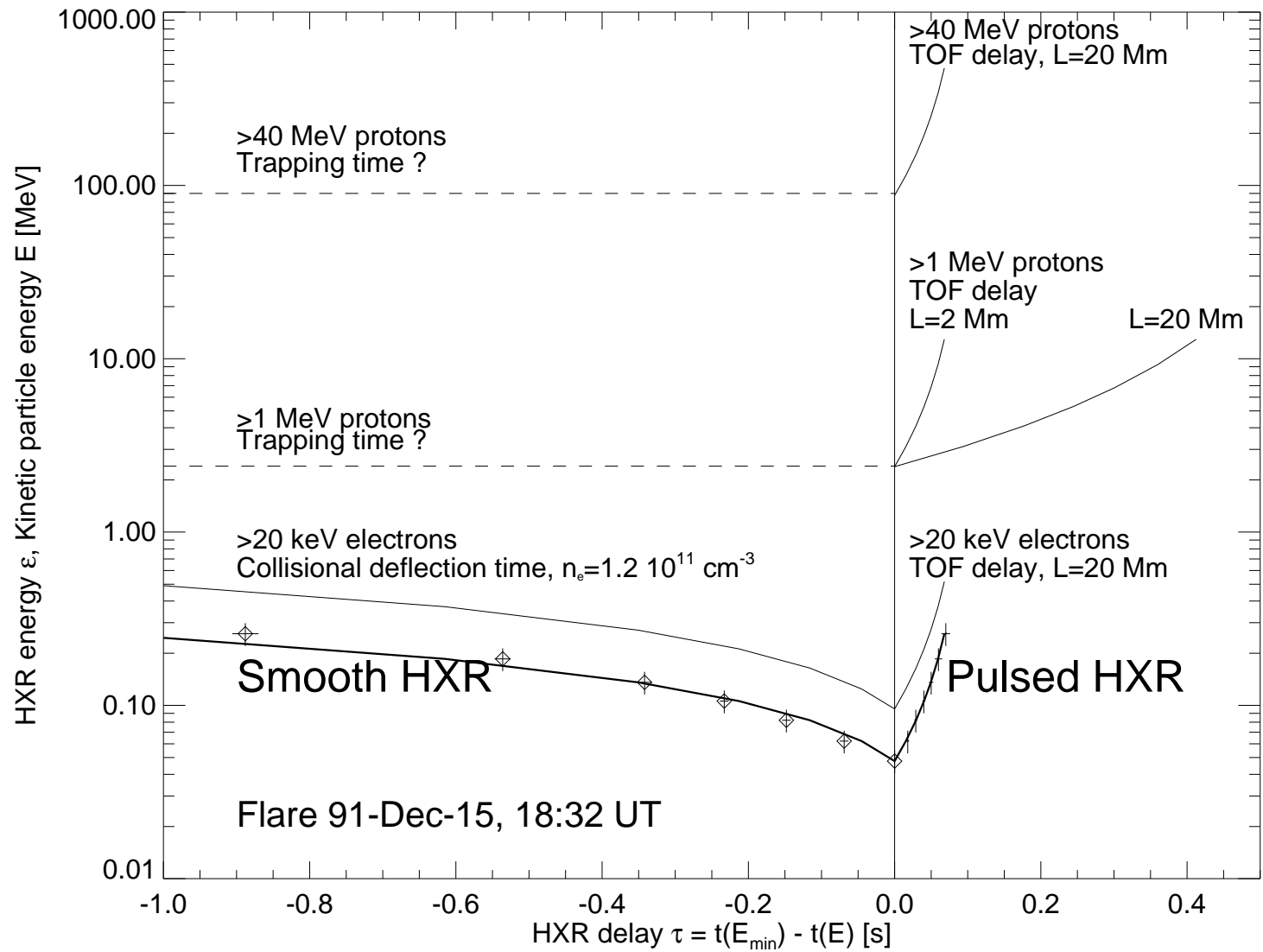
\begin{figure}
\caption{The energy-dependent time delays $\tau(E)=t(40$ keV$)-t(E)$
are shown for a flare observed on 1991 Dec 15, separately measured for the 
the {\sl pulsed} HXR flux (crosses on right-hand side) and the {\sl smooth} 
HXR flux (diamonds on left-hand side), in the energy range of 40-300 keV. 
The delays can be fitted with two models: in terms of electron time-of-flight 
(TOF) differences (thick line on right-hand side bottom), and in terms of
electron collisional deflection time differences (thick line on left-hand side
bottom). The electron energies ($E \approx 2 \varepsilon$) with the 
same timing delays as the HXR pulses 
(observed at energy $\varepsilon$) are indicated
with thin lines (bottom part). For comparison, we show also the required timing
of $>1$ MeV protons (left-hand side middle) 
and $>40$ MeV protons (left-hand side
top), if they would be responsible for the $>20$ keV HXR emission. 
The trapping time of $>1$ MeV protons (dashed line on left-hand side middle)
or $>40$ MeV protons (dashed line on left-hand side top) are so large 
(estimated with Eqn.6) that no energy-dependence can be seen on the displayed 
time scale of 1 s.}
\end{figure}
\end{document}